\documentclass[a4paper]{PoS}

\usepackage{url}

\title{Sun/Moon photometer for the Cherenkov Telescope Array - first results}

\ShortTitle{Sun/Moon photometer for CTA - first results}

\author{\speaker{Jakub Jury\v{s}ek}$^1$ and Michael Prouza$^1$ for the CTA Consortium\\
\llap{$^1$}Institute of Physics, The Czech Academy of Sciences, Prague, Czech Republic\\
E-mail: \email{jurysek@fzu.cz}}

\abstract{
Determination of the energy and flux of the gamma photons by Imaging Atmospheric Cherenkov Technique is strongly dependent 
on optical properties of the atmosphere. Therefore, atmospheric monitoring during the future observations of the Cherenkov 
Telescope Array (CTA) as well as anticipated long-term monitoring in order to characterize overal properties and annual
variation of atmospheric conditions are very important. Several instruments are already installed at the CTA sites in order to 
monitor atmospheric conditions on long-term. One of them is a Sun/Moon photometer CE318-T, installed at the Southern CTA site. 
Since the photometer is installed at a place with very stable atmospheric conditions, it can be also used for characterization
of its performance and testing of new methods of aerosol optical depth (AOD) retrieval, cloud-screening and calibration. In this work,
we describe our calibration method for nocturnal measurements and the modification of cloud-screening for purposes of nocturnal 
AOD retrieval. We applied these methods on two months of observations and present the distribution of AODs in four photometric 
passbands together with their uncertainties.}

\FullConference{35th International Cosmic Ray Conference - ICRC2017-\\
		10-20 July, 2017\\
		Bexco, Busan, Korea}

\begin{document}

\section{Introduction}

The Sun/Moon photometer CE318-T, installed at the Southern site of the Cherenkov Telescope Array (CTA) in Chile, is part of 
the auxiliary scientific instrumentation developed for the atmosphere monitoring and calibration of the CTA \cite{ebr2017}. The photometer
can measure day-time and night-time evolution of the integral optical depth of the atmosphere with very high 
precision in 9 photometric passbands (with the central wavelengths at 340, 380, 440, 500, 675, 870, 937, 1020 and 1640 nm) with high cadence of one measurement
every three minutes. This allows an understanding of the detailed characteristics of the atmospheric extinction and the 
underlying particulates, mainly aerosols.

The usual disadvantage of passive devices measuring AOD at night is the low incoming light flux from stars. Using the 
Moon as a light source with brightness $10^5$-times bigger than that of the brightest star Sirius enables the photometer 
to reach a high signal-to-noise ratio and therefore also a precision unreachable for star photometers. The only disadvantage 
is that the photometer does not measure continuously at night but requires at least 40 \% illumination of the Moon and 
its altitude above 10 degrees in order to obtain reasonable precision of all measurements. 

The photometer is a completely stand-alone device, equipped with built-in batteries, solar panel, GPS synchronization and a control
unit which enables autonomous measurements and data acquisition. It was installed at the Southern site of the CTA in June 2016 and
since then it collected data until September 2016. In this work we present the analysis of the last two months of 
observations.

Installed at a high altitude of 2154 metres above sea level, near Cerro Paranal, a place with very stable atmospheric conditions, 
the photometer can also be used for the characterization of its performance and for testing new methods of AOD retrieval, 
cloud-screening and calibration. This is very important in the case of Moon photometry, which is still not completely 
optimized. In this work, we discuss our calibration method for nocturnal measurements and modification of usual cloud-screening 
for the purposes of the Moon photometry.

\section{Calibration} \label{calibration}

The photometer is part of AERONET (AErosol RObotic NETwork)\footnote{https://aeronet.gsfc.nasa.gov/}, a wordwide network of precisely calibrated instruments, which
provides globally distributed observations of diurnal AODs. It was calibrated for diurnal measurements in AERONET in 
November 2016, which allows us to compare the results of our on-site calibration with those obtained by AERONET and to
estimate systematic uncertainties of our AODs. However, photometry of the
Moon and calibration of nocturnal measurements is still not satisfactorily understood and AERONET does not yet provide
calibration of the photometer for the purposes of nocturnal measurements. In order to obtain night-time AODs we had to
develop our own methods based on recently published papers. 

For the purpose of diurnal calibration, the classical Langley method was used. The Langley method is based on a simple 
linear fit of the equation (\ref{eq.langley}) based on the Lambert-Beer's law
\begin{equation} \label{eq.langley}
\ln V = \ln (V_0 / R^2) - X \tau
\end{equation}
where $V$ stands for measured voltage, $X$ for airmass, 
$\tau$ for optical depth, $R$ for Sun-Earth distance and $V_0$ is 
the so-called extraterrestrial voltage, assumed constant in a given photometric passband. According to the suggestion made by Shaw \cite{shaw1983}, 
only morning measurements on a short interval of airmasses from 2 to 5 were used, because of more stable atmospheric 
conditions in the morning than after noon. After a preliminary cloud-screening based on the AERONET V3 algorithm 
(see further details in section \ref{cloudscreening}) sigma-clipping was applied in order to reject outliers
and remaining measurements affected by clouds. Finally, only those fits with empirically estimated threshold on $RMS$ of the distribution of residuals as $\mathrm{RMS} \leq 0.01$ were 
kept and the final precision of the mean values of $V_0$ for each passband\footnote{The 937 nm passband is not included, because there is 
strong contribution of water vapour to absorbtion and the total optical depth is therefore highly varible.}
 is listed in Table \ref{tab.calibration}, together with its comparison to those obtained from AERONET.

\begin{table}
\centering
\begin{minipage}{.48\textwidth}
\begin{tabular}{ccc}
\hline
$\lambda$ &	$\sigma_{V0\, \rm{CTA}}$  & $(V_{0\, \rm{CTA}} - V_{0\, \rm{AERO}})/ V_{0\, \rm{AERO}}$ \\
(nm) & (\%) & (\%) \\
\hline
340          &		1.51  &	1.15	 	\\
380    	 &		1.38     &  	1.92	\\
440     	 &		1.17 & 	0.66   	\\
500    	 &		1.05     &  	0.92	\\
675     	 &		0.76  &  	0.41  	\\
870     	 &		0.64  &  	0.34  	\\
\hline
\end{tabular}
\caption{Precision of calibration constants for diurnal observation	and comparison with AERONET.}
\label{tab.calibration}
\end{minipage}\hfill
\begin{minipage}{.48\textwidth}
\begin{tabular}{ccc}
\hline
$\lambda$ &	 $\sigma_\kappa(|g| < \pi/4)$ &	$\sigma_\kappa(|g| > \pi/4)$ \\
(nm) & (\%) & (\%) \\
\hline
340		&	--		&		--		\\
380		&	--		&		--		\\
440     & 	0.63	 &		1.77    \\
500     &  	0.41	 &		0.97   	\\
675     &  	0.26	 &		0.61   	\\
870     &  	0.28	 &		0.55   	\\
\hline
\end{tabular}
\caption{Precision of calibration constants for nocturnal observation for different Lunar phases ($g$).}
\label{tab.kappa_prec}
\end{minipage}\hfill
\end{table}

The case of calibration of nocturnal measurements is much more difficult, because illumination of the Moon varies with time
and therefore also the extraterrestrial voltage strongly depends on the phase of the Moon and the standard Langley method has to be 
modified. Barreto et al. \cite{barreto2013} developed the so-called Lunar-Langley method, where the extraterrestrial voltage 
$V_0$ in each passband is modified as $V_0 = I_0 \kappa$, where $\kappa$ is a calibration constant and $I_0$ is an
extraterrestrial irradiance of the detector, which can be calculated using the ROLO lunar reflectance model \cite{rolo2005}.
The lunar reflectance in the ROLO model is given as a complicated empirically derived analytical function of geometrical variables
like the absolute value of the phase angle, the selenographic longitude of the Sun and selenographic latitude and longitude of the observer, respectively.
These variables have to be obtained with the use of some reliable source of precise ephemerides. In this 
work, we used JPL (Jet Propulsion Laboratory) Horizons ephemerides obtained via the \texttt{Python} library \texttt{callhorizons}.\footnote{Available at \url{https://pypi.python.org/pypi/CALLHORIZONS}} 

\begin{figure}[!t]
\centering
\begin{tabular}{cc}
\includegraphics[width=.49\textwidth]{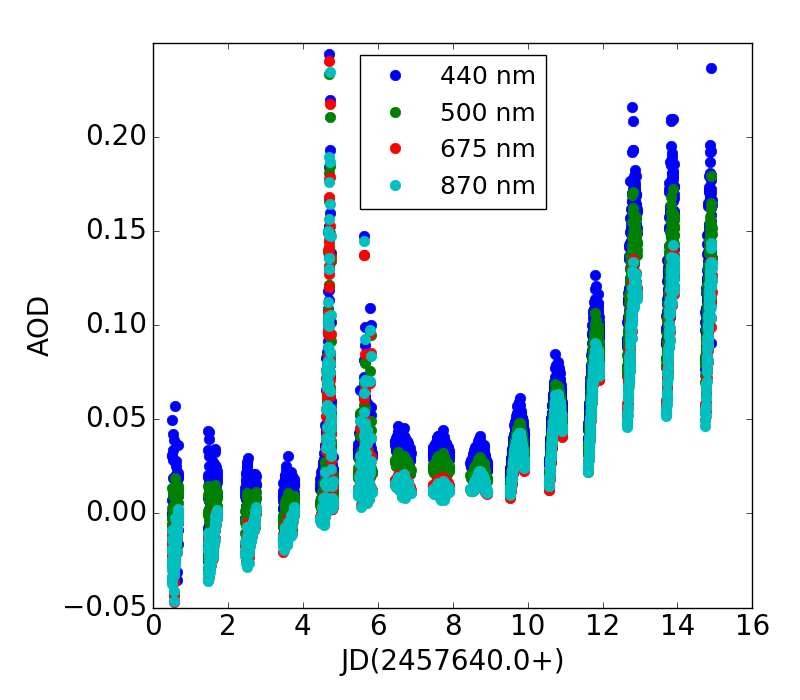} & \includegraphics[width=.49\textwidth]{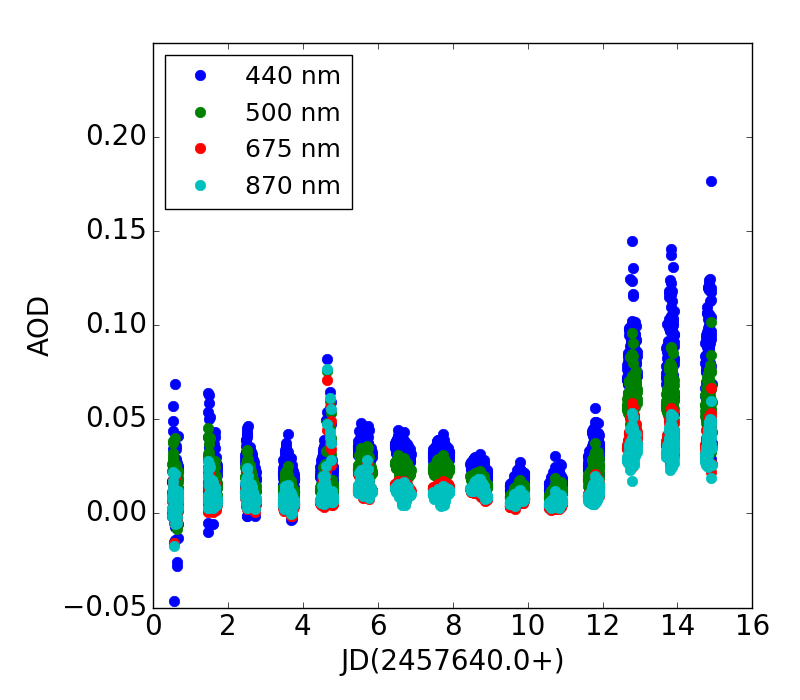} \\
\end{tabular}
\caption{\textit{Left:} Nocturnal AOD in four passbands with the use of one value of $\kappa$ for all observations. \textit{Right:} Nocturnal 
AOD corrected for the dependency of $\kappa$ on the phase of the Moon. (JD = Julian date time format)}
\label{fig.aods}
\end{figure}

\begin{figure}[!t]
\centering
\begin{tabular}{cc}
\includegraphics[width=.49\textwidth]{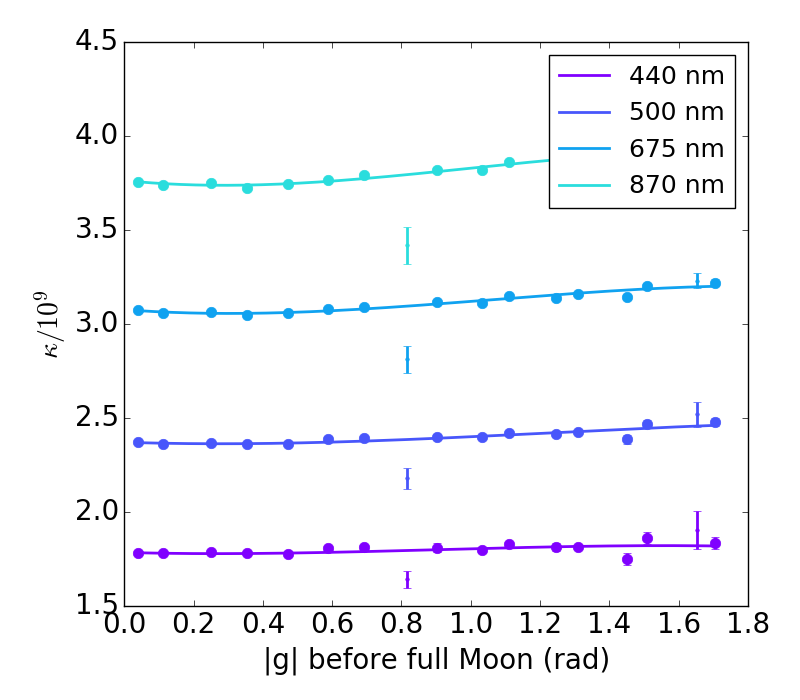} & \includegraphics[width=.49\textwidth]{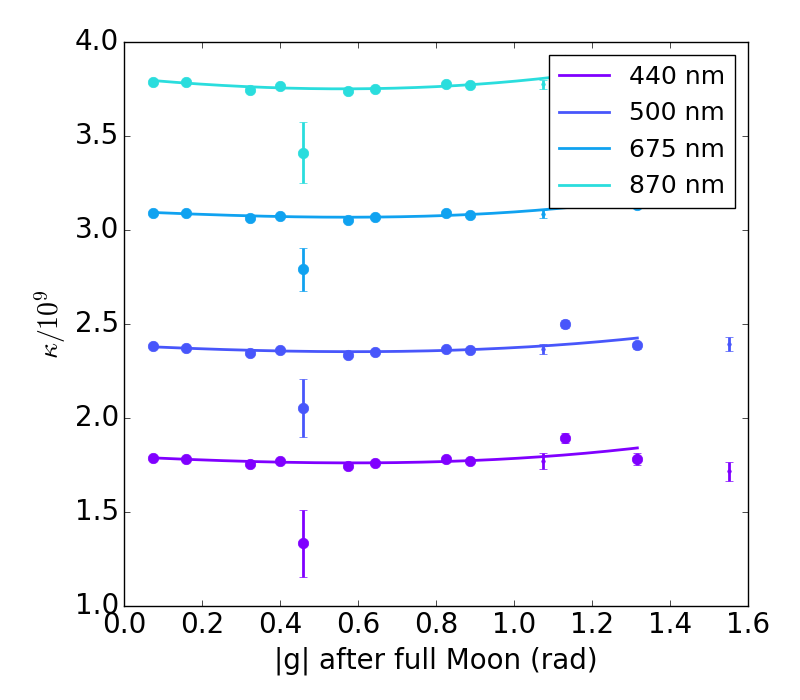}  \\
\end{tabular}
\caption{Phase dependence of $\kappa$.}
\label{fig.kappa_phase}
\end{figure}

The accuracy of the ROLO model plays a critical role in the calibration. The authors of the original paper quoted an accuracy 
of better than $1 \%$. However, during our analysis it turned out that after Lunar-Langley calibration,
we obtained a lot of negative AODs or apparent nocturnal cycles of AODs (see left Fig. \ref{fig.aods}),
occuring when the calibration is wrong (e.g. \cite{cachorro2004}). This means that $\kappa$ is not really a constant and the extraterrestrial voltage,
although corrected by the ROLO model, still varies with time. 
This problem was independently found by Barreto et al. \cite{barreto2016} and they also
recently proposed an original method to overcome it \cite{barreto2017}. They kept the $\kappa$ constant and fitted
differences between extrapolated diurnal AODs and measured nocturnal AODs with a polynomial function of the Moon phase, 
modulated by airmass. Our approach was different. Since we have a lot of Langley plots from many clear nights with 
stable conditions, we were able to study possible dependencies of $\kappa$ and it turned out that $\kappa$ strongly 
depends on the phase of the Moon as shown in the Fig. \ref{fig.kappa_phase}. We therefore fitted the dependence with a third order polynomial to
 obtain corrected $\kappa$ for each Moon phase. An example of corrected nocturnal AODs is shown in the
right panel of Fig. \ref{fig.aods} and it can be seen that there are no residual nocturnal cycles. 
RMS values the of residuals of the fits of the $\kappa$ dependence on the Moon phase are listed in Tab. \ref{tab.kappa_prec}.

The persistent question is whether the ROLO model is accurate enough, and if not, why. This would be, of course, one of the possible explanations for the residual discrepancy. 
But as Tom Stone pointed out \cite{stone_talk2017}, also temperature variations of the sensor, or non-linearities of the 
detector on low fluxes from nocturnal observation taken far from the full Moon phase can cause the observed phase 
dependence of $\kappa$. 
It is still an open question and matter of intensive research.

\section{Cloud-screening}\label{cloudscreening}

A key part of the AOD retrieval is cloud-screening, because even optically thin clouds in the path of observation strongly 
affect the obtained value of AOD. 
In this work, we used a method based on a modified AERONET V3 algorithm 
\cite{smirnov2000,holben_talk2013} consisting of two steps - applying the triplet stability criterion and a smoothness check. The AERONET
algorithm was developed for diurnal measurements only and the thresholds were universally set on large data samples from many observing 
sites. In the next paragraphs we describe our readjustment of the method to our site and its modification for the purpose of 
nocturnal measurements.

\begin{figure}[!t]
\centering
\begin{tabular}{cc}
\includegraphics[width=.49\textwidth]{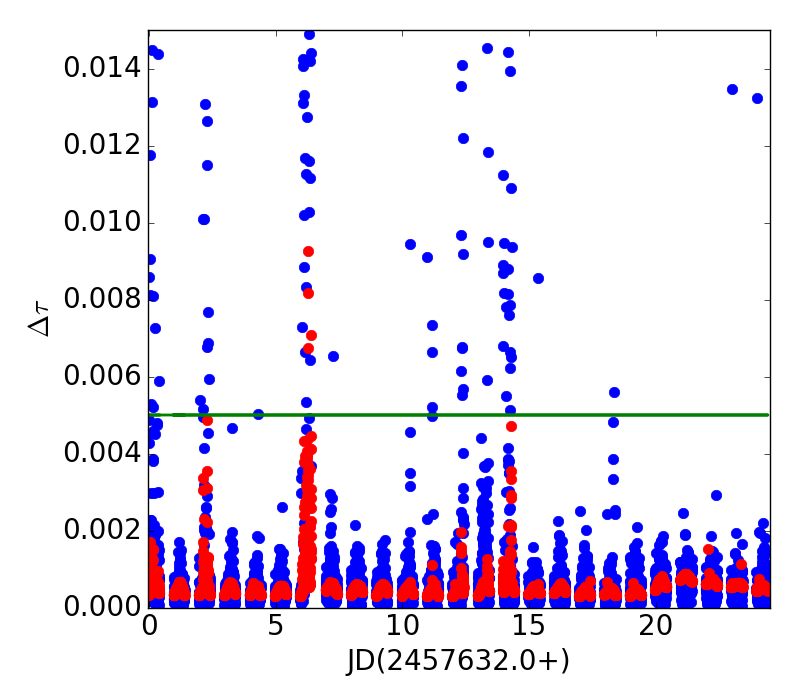} & \includegraphics[width=.49\textwidth]{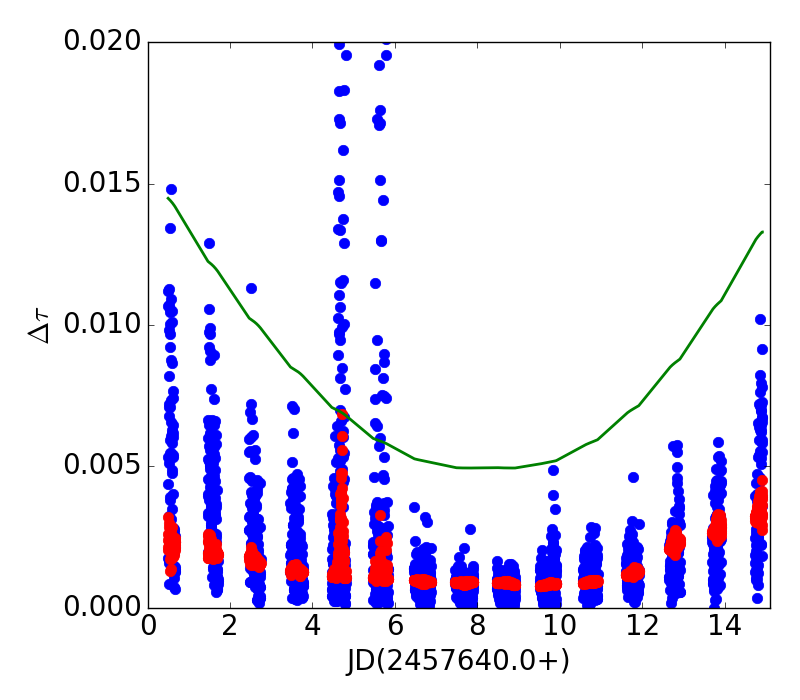}  \\
\end{tabular}
\caption{\textit{Left:} Triplet stability for diurnal measurements. The green line represents the 0.005 threshold, the $0.015\tau$ threshold is marked as a red dot for each observation. \textit{Right:} Triplet stability for nocturnal measurements.}
\label{fig.triplets}
\end{figure}

The triplet stability criterion for diurnal observations according to the AERONET V3 algorithm states that a given measurement is 
affected by clouds if $\Delta \tau > \mathrm{max}(0.01,0.015 \tau)$ in all 675, 870 and 1020~nm passbands, where 
$\Delta \tau$ is the difference between the maximal and minimal value of the triplet. However, it turned out that this 
general criterion is too loose in the case of our data, taken at the site under stable atmospheric conditions. We 
therefore decided to lower the 0.01 threshold to 0.005 based on empirical observation. An example of triplet stability of 
our diurnal measurements is shown in the left panel of Fig. \ref{fig.triplets} for one month of observation. 

In case of lunar measurements, the situation becomes more difficult. Since the triplet stability depends on the incident 
flux, it also depends on the lunar phase and the criterion has to be modified. In first approximation, the dependence
can be described as a simple polynomial of second order as shown in the right panel of Fig. \ref{fig.triplets}. Based on our empirical assessment, we modified the criterion
like 
\begin{equation}
\Delta \tau < \mathrm{max}[(0.005,0.015 \tau) \mathrm{P}(g) ],
\end{equation}
where $\mathrm{P}(g)$ is a function of the Moon phase $g$ given by equation
\begin{equation} \label{pg}
\mathrm{P}(g) = 0.8g^2 - 0.2|g| + 1.
\end{equation}

\begin{figure}[!t]
\centering
\begin{tabular}{cc}
\includegraphics[width=.49\textwidth]{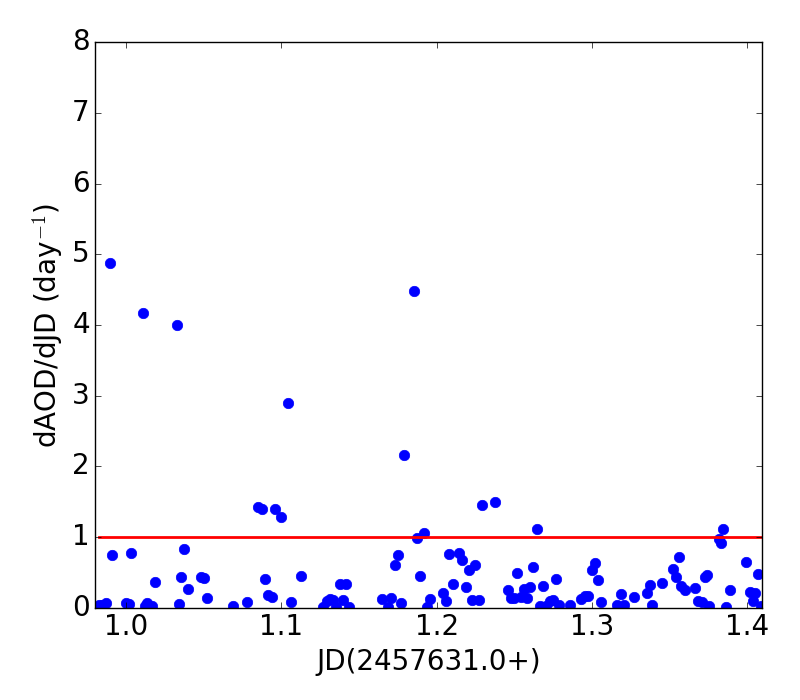} & \includegraphics[width=.49\textwidth]{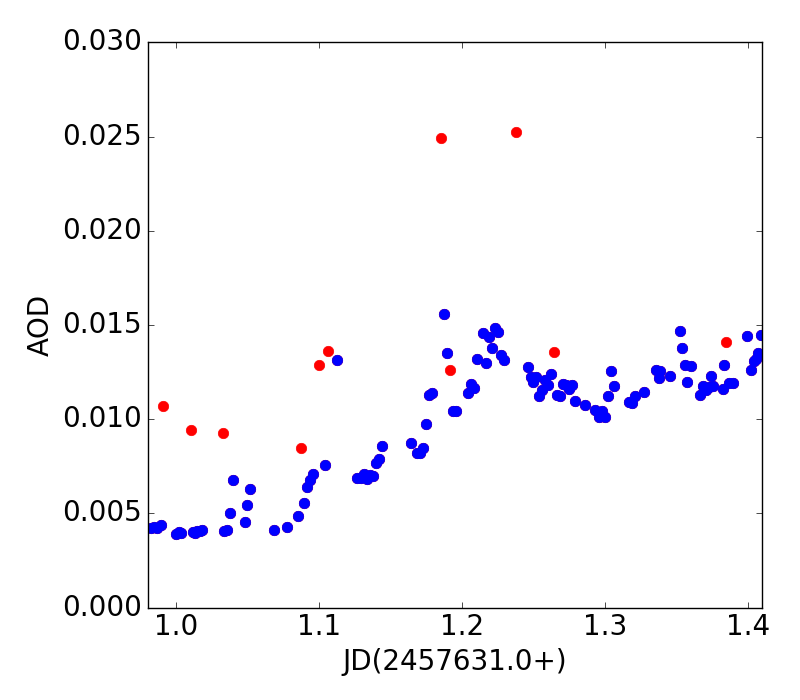}  \\
\includegraphics[width=.49\textwidth]{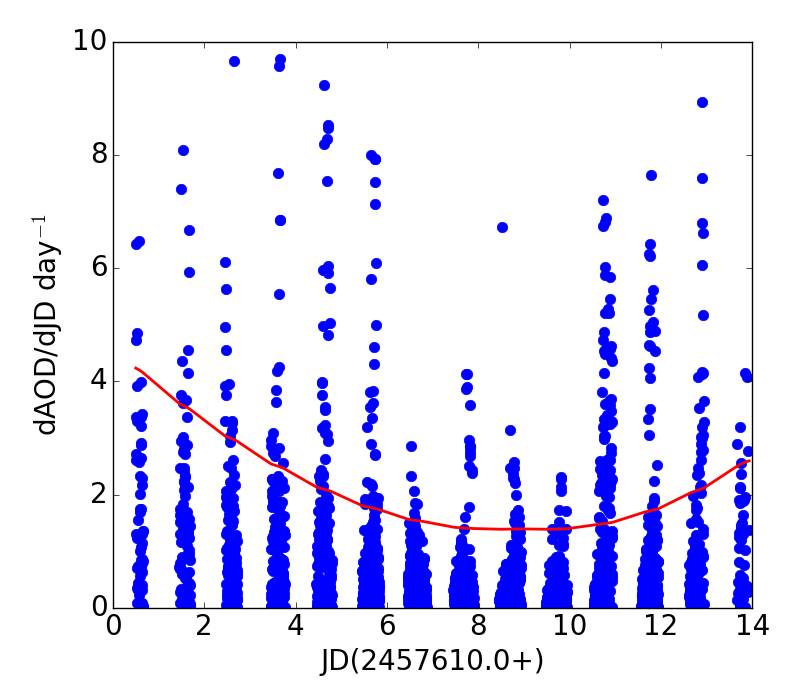} & \includegraphics[width=.49\textwidth]{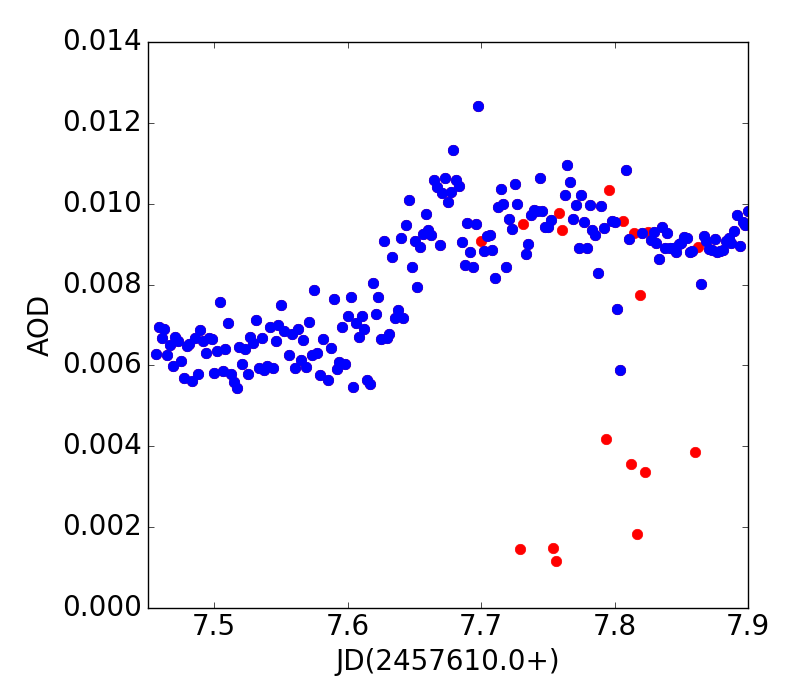}  \\
\end{tabular}
\caption{\textit{Upper and bottom left:} Smoothness check for diurnal (top) and nocturnal (bottom) measurement. Empirical thresholds are marked 
as red lines. \textit{Upper and bottom right:} Examples of AODs after the smoothness check. Rejected measurements are marked as red dots.}
\label{fig.smoothness}
\end{figure}

The second step of the cloud-screening, the smoothness check, assumes that the AODs do not vary too rapidly. According to the AERONET V3
algorithm it means that, in case of diurnal observations, $\Delta \mathrm{AOD}/ \Delta \mathrm{JD}$\footnote{JD is so-called Julian date.} is required to be less than $14.4 \, \mathrm{days^{-1}}$ 
for each pair of measurements in the 500 nm passband if there are no clouds. As shown in Fig. \ref{fig.smoothness}, in our 
case the general criterion is also too loose and we had to lower the threshold to the empirical value of $1 \, \mathrm{day^{-1}}$.
This lead to the rejection of most outliers, which we believed were caused by clouds. 

For the purpose of lunar measurements the criterion had to be modified, due to the fact that 
$\Delta \mathrm{AOD}/ \Delta \mathrm{JD}$ is phase dependent as shown in Fig. \ref{fig.smoothness}. We used the same 
function as (\ref{pg}) to modulate the diurnal criterion like $\Delta \mathrm{AOD}/ \Delta \mathrm{JD} < 1.4 \mathrm{P}(g) \, \mathrm{days^{-1}}$.

\section{Results}

The methods described above were applied on data taken during two months of observation -- August and September 2016.
Final values of AODs were obtained after subtraction of molecular absorption computed with the use of a MODTRAN model \cite{modtran}. 
Since the flux of the moonlight is too low for shorter wavelength, the lowest two channels -- 340 and 380 nm -- were not used in the 
analysis. In addition, our MODTRAN model covers wavelengths only up to 900 nm and therefore channels 937, 1020 and 1640 were also excluded from 
the processing.

\begin{figure}[!t]
\centering
\begin{tabular}{cc}
\includegraphics[width=.48\textwidth]{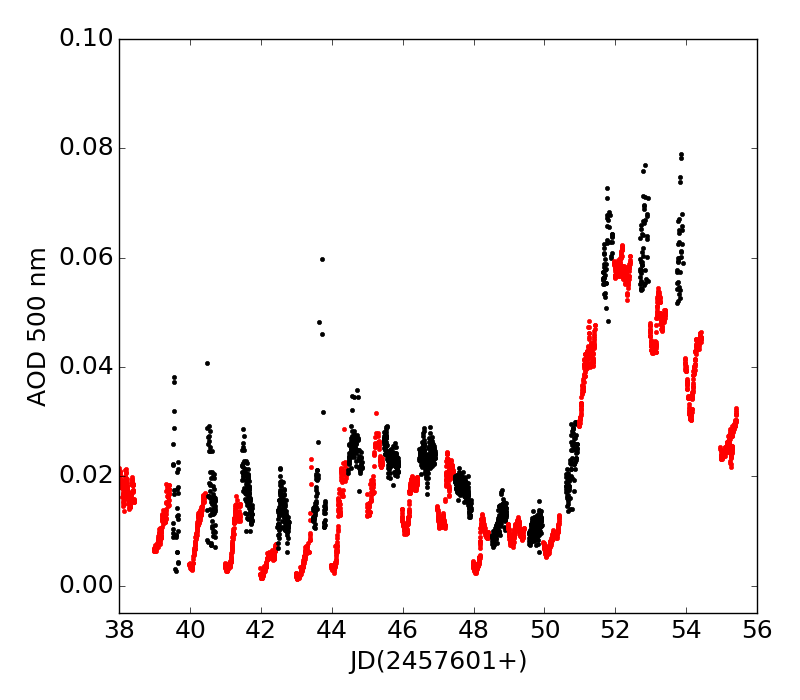} & \includegraphics[width=.48\textwidth]{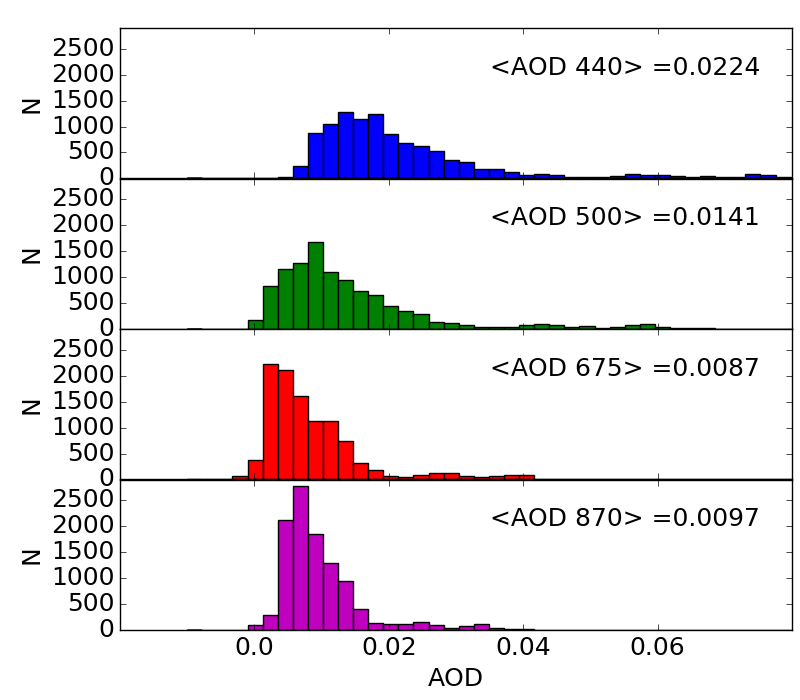}  \\
\end{tabular}
\caption{\textit{Left:} Example of the time dependence of AODs in the 500 nm passband for diurnal (red) and nocturnal (black) 
measurements combined. \textit{Right:} overall distribution of AODs in four photometric passbands.}
\label{fig.vaods}
\end{figure}

Diurnal AODs together with nocturnal ones for the 500 nm passband for one month of observation are shown in Fig. \ref{fig.vaods}. 
Despite the fact that AODs show rapid variation,
nocturnal measurements connect very well to the diurnal measurements. In some cases the nocturnal AODs are higher 
than expected from the diurnal trend (for example days 53 and 54 from the beginning of the time interval).We think that this is most probably caused 
by imperfect calibration. Note that the calibration error of $1\,\%$ results in an uncertainty in the optical depth of $\approx0.01$ for
optical airmass $X=1$ if other sources of uncertainties are neglected. 
The calibration accuracy naturally decreases with decreasing lunar 
irradiance and is therefore phase dependent as shown in Table \ref{tab.kappa_prec}. Thus a mismatch of the order of $\approx0.01$ 
in AOD is expected for phases far from the full moon. 

One can also notice that in most cases AOD increases during the day. At this point, we are not sure whether these
trends are real or not. This effect is known for example from Mauna Loa observations, where AODs increase in the afternoon
due to the upslope winds. But we have to be careful, because such behavior might also be related to the unstable temperature of the
detector. 

\begin{figure}[!t]
\centering
\begin{tabular}{cc}
\includegraphics[width=.48\textwidth]{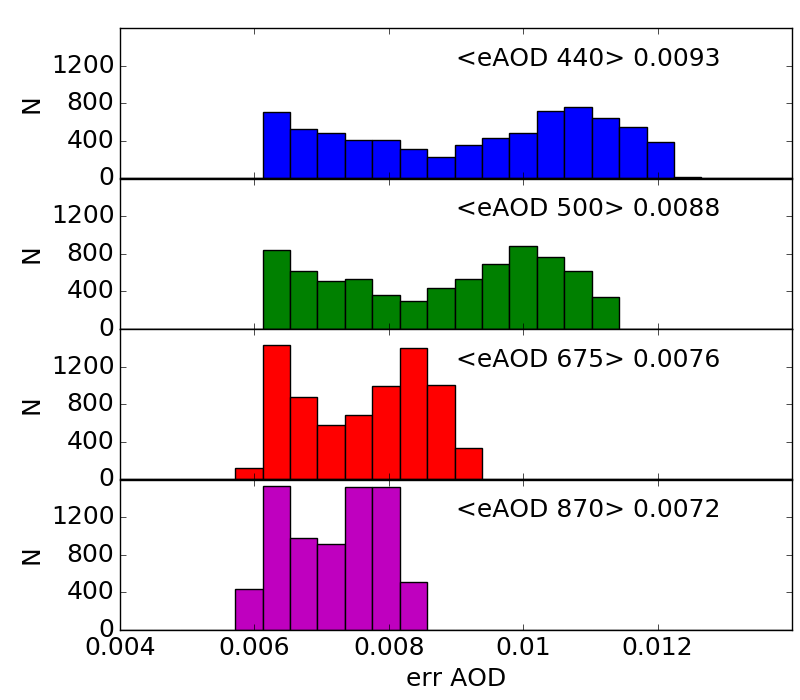}  & \includegraphics[width=.48\textwidth]{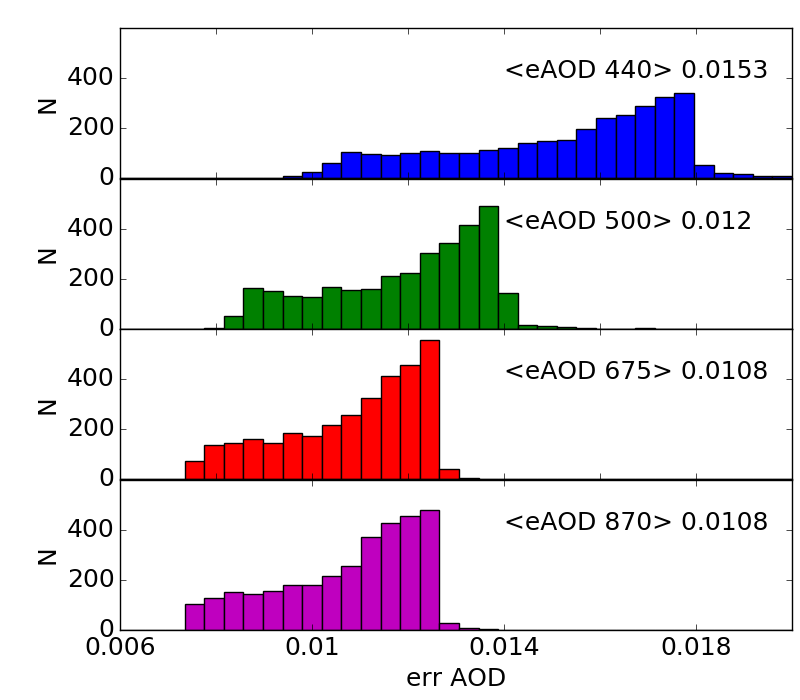} \\
\end{tabular}
\caption{Distribution of uncertainties of diurnal (left) and nocturnal (right) measurements.}
\label{fig.evaod_hist}
\end{figure}

The overall statistics of AODs retrieved at the southern CTA site is shown in the right panel of Fig. \ref{fig.vaods} and 
the distribution of uncertainties is shown in Fig. \ref{fig.evaod_hist}. Uncertainties of diurnal AODs 
($u^\mathrm{D}_\mathrm{AOD}$) and nocturnal AODs ($u^\mathrm{N}_\mathrm{AOD}$), respectively, were calculated from the following relations
\begin{equation}
(u^\mathrm{D}_\mathrm{AOD})^2 = \frac{1}{X^2}\frac{u(V_0)^2}{V_0^2} + u_\mathrm{sys}^2, \;\;\;\;\; (u^\mathrm{N}_\mathrm{AOD})^2 = \frac{1}{X^2}\biggl(\frac{u(\kappa)^2}{\kappa^2} + \frac{u(I_0)^2}{I_0^2} + \frac{u(V)^2}{V^2} \biggr) + u_\mathrm{sys}^2,
\end{equation}
where $u(V_0)/V_0$ and $u(\kappa)/\kappa$ stand for relative accuracy of the calibration constants, $u(I_0)/I_0$ is the relative 
accuracy of the ROLO model ($\approx$ 1 \%) and $u(V)/V$ is relative precision of each measurement, derived from triplet 
variability \cite{barreto2016}. Systematic uncertainties ($u_\mathrm{sys}$) were estimated from differences between our and the AERONET
calibration.

\acknowledgments{This work was conducted in the context of the CTA CCF Work Package. 
We gratefully acknowledge financial support from the agencies and organizations listed here: http://www.cta-observatory.org/consortium\_acknowledgments.
This work was also supported by the Czech Ministry of Education of 
the Czech Republic (MSMT LTT17006, LM2015046 and  CZ.02.1.01/0.0/0.0/ /16\_013/0001403).}

\end{document}